# PATH INTEGRAL REPRESENTATION OF COMPOSITE FERMIONS AND BOSONS

P. Béran
*ATM-Technology Center, 3965 Chippis, Switzerland*

The density matrix of the 2D system of spinless electrons confined to the lowest Landau level is expressed using both basis of states parametrized by electron locations and basis of states parametrized by hole locations. In this representation, the electron-electron repulsion can be viewed as an electron-hole *attraction*. Electron-hole pairs stabilized by this attraction provide a new formulation for Composite Fermions *which fully respects particle-hole symmetry*. This representation also allows a particularly simple formulation of the composite Boson approach of generic *v=p/q* incompressible states: The *v=p/q* state corresponds to the formation of clusters made up on *p* electrons and *q-p* holes and fractionally charged excitations correspond to the breaking of such clusters.

## 1. INTRODUCTION

It is now admitted that the composite Fermion approach [1,2] provides an adequate description of incompressible and compressible Fractional Quantum Hall states. Both composite Fermion approach and composite Boson picture [3,4] of incompressible states were originally formulated using a singular gauge transformation, which maps the system of electrons onto a system of composite particles consisting in electrons attached to an even number of flux quanta (composite Fermions) or to an odd number of flux quanta (composite Bosons). Latter on, in formulating a microscopic description of the *v=1/2* compressible state, Read proposed to describe Composite Fermions as electron-hole pairs [5]. However, none of these approaches preserved the particle-hole symmetry present in the electron system confined to the lowest Landau level. Recently, there has been an increasing interest in formulating a description of Composite Fermions which preserves this symmetry [6,7].

In this note, an alternative formulation of Composite Fermions is presented which explicitly preserves the particle-hole symmetry. This formulation makes use of a particular representation of the finite temperature density matrix for the spinless electron system confined to the lowest Landau level which consists in matrix elements of the density matrix operator evaluated between states parametrized by electron coordinates and states parametrized by the coordinates of electron vacancies, i.e. holes [8,9]. In this representation, the electron-electron repulsion can be viewed as an electron-hole *attraction*. In presence of this attraction and for filling factor close to *1/2*, we postulate that, for some temperature, the density matrix takes significant values only for configurations of electron coordinates and hole coordinates in which coordinates can be grouped into clusters made up of one electron coordinate and one hole coordinate each. The partition function of the electron system is then expressed as an integral over paths followed by the guiding centers of these clusters in imaginary time. We find that the phase associated to these paths equals to that expected for composite Fermions and we thus identify these clusters with composite Fermions.

We also make the case that the density matrix formalism developed here allows a direct formulation of the composite Boson approach of incompressible states for filling factors *v=p/q* with arbitrary numerator *p≥1*, in contrast to the formulation based on singular gauge transformation in which *v=p/q* states are constructed iteratively. In this formulation, an incompressible state corresponds to the formation of cluster made up of *p* electrons and *h=q-p* holes in the density matrix. The partition function of the electron system is expressed as an integral over paths followed by the guiding centers of such clusters in imaginary time and it is found that if the product *ph* is even, these paths contribute constructively to the partition function, leading to superfluidity at low temperature. This formulation of the composite Boson approach presents several conceptual advantages: (a) the electron-hole symmetry is explicitly preserved, (b) the absence of incompressible states $v=p/q$ with even denominator in the lowest Landau level can be simply and generically explained in terms of cluster stability, (3) quasiparticle excitations can be generically described as corresponding to the breaking of a cluster made up of $p$ electrons and $h$ holes in two smaller clusters, each one corresponding to a defect with fractional charge.

## 2. COMPOSITE FERMIONS

Let us first study the density matrix of the 2D gas of spinless electrons confined to the lowest Landau level at filling factor $v=p/q$. More specifically, we consider the quantity

$$\rho_\beta(w_1,...,w_{hN},z_1,...,z_{pN}) = \langle w_1...w_{hN} | e^{-\beta \bar{V}} | z_1...z_{pN} \rangle \tag{1}$$

where $\bar{V}$ is the electron-electron repulsion projected onto the lowest Landau level and where $|z_1...z_{pN}\rangle$ and $|w_1...w_{hN}\rangle$ are states of the electron system with filling factor $v=p/q$, respectively given in electron representation and hole representations, defined by

$$|z_1...z_{pN}\rangle = c^+_{z_1}...c^+_{z_{pN}}|0\rangle \tag{2}$$

and by

$$|w_1...w_{hN}\rangle = c_{w_{hN}}...c_{w_1}|1,qN\rangle \tag{3}$$

where $|0\rangle$ and $|1,qN\rangle$ respectively denote the empty state and the full-Landau level state with $qN$ electrons, $c_w$ is the operator annihilating an electron in a coherent state [10] centered at position $w$, $h$ is given by $h=q-p$ and where $pN$ is the number of electrons in the lowest Landau level.

For $\beta=0$, the density matrix of Eq. (1) is simply given by

$$\rho_0 = \prod_{1 \leq j_1 < j_2 \leq hN}(w_{j_1}-w_{j_2}) \prod_{1 \leq i_1 < i_2 \leq pN}(z_{i_1}-z_{i_2}) \prod_{j=1}^{hN}\prod_{i=1}^{pN}(w_j-z_i) \prod_{j=1}^{hN} e^{-\frac{|w_j|^2}{4}} \prod_{i=1}^{pN} e^{-\frac{|z_i|^2}{4}} \tag{4}$$

in units of the magnetic length, where $z_i$ and $w_j$ are respectively the complex coordinates of electron $i$ and hole $j$.

Let us now consider the effect of the electron-electron repulsion present in Eq. (1). When projected onto the lowest Landau level, this repulsion takes the form

$$\bar{V} = \int d^2z_1 d^2z_2 d^2z'_1 d^2z'_2 \sum_{n=0}^{\infty} \tilde{V}_{2n+1} K_{2n+1}(z_1,z_2,z'_1,z'_2) c^+_{z'_1} c^+_{z'_2} c_{z_2} c_{z_1} \quad, \tag{5}$$

where $\tilde{V}_L$ denotes pseudo-potential coefficients and where

$$K_L(z_1,z_2,z'_1,z'_2) = \pi^{-1}\int d^2\zeta \, \psi_L^{(\zeta)*}(z_1,z_2) \, \psi_L^{(\zeta)}(z'_1,z'_2) \tag{6}$$

is a kernel used for projecting onto two-electrons states with definite relative angular momentum $L$, given in terms of the wave function

$$\psi_L^{(\zeta)}(z_1,z_2) = \frac{1}{2^{L+1}\pi\sqrt{L!}}(z_1^*-z_2^*)^L e^{\frac{z_1^*+z_2^*}{2}\zeta - \frac{|z_1|^2+|z_2|^2}{4} - \frac{|\zeta|^2}{2}} \tag{7}$$

which describes a pair of electrons with center of mass in a coherent state centered at position $\zeta$ and with definite relative angular momentum $L$. Using commutation relations of creation and annihilation operators in Eq. (5) in order to move $c_{z_1}$ to the left of $c^+_{z'_1}$ and exploiting the invariance of function $K_L$ under global translation of its variables, the matrix elements of the electron-electron repulsion can be written as

$$\langle w_1...w_{hN}|\overline{V}|z_1...z_{pN}\rangle = -\sum_{\substack{i=1...pN \\ j=1...hN}}\sum_{n=0}^{\infty}\tilde{V}_{2n+1}\left[P_{(w_j,z_i)}^{(L)}\rho_0\right](w_1,...,w_{hN},z_1,...,z_{pN}) + cte,$$

(8)

where *cte* denotes a term which is independent of coordinates *w*'s and *z*'s and where $P_{(w_j,z_i)}^{(L)}$ is the operator projecting onto two-electrons states with definite relative angular momentum *L*, defined by

$$\left[P_{(w_j,z_i)}^{(L)}\rho_0\right](w_1,...,w_{hN},z_1,...,z_{pN}) =$$
$$\int d^2z_i' d^2w_j'\, K_L(w_j,z_i,w_j',z_i')\, \rho_0(w_1,...,w_j',...,w_{hN},z_1,...,z_i',...,z_{pN}) \,.$$

(9)

The right-hand side of Eq. (8) can be interpreted as the energy of a system of particle with coordinates $w_1,...,w_{hN},z_1,...,z_{pN}$ in which an *attractive* interaction $-\overline{V}$ is switched on for pairs $(w_j,z_i)$ only. Using the fact that the left-hand side of Eq. (8) equals $-\partial\rho/\partial\beta|_{\beta=0}$, it can be concluded that the effect of a short range repulsion is to increase the value of the density matrix $\rho_\beta$ for configurations $(w_1,...,w_{hN},z_1,...,z_{pN})$ in which a hole coordinate $w_j$ can be found in the vicinity of every electron coordinate $z_i$, i.e. at a distance of the order of the magnetic length.

We now restrict to the case where ν is slightly below ½ and attempt to write down an ansatz for the density matrix satisfying the latter requirement. More specifically, we assume that, for some value $\beta_0$ of the inverse temperature, the density matrix can be approximated by

$$\rho_{\beta_0} \propto A_w A_z \left[P_{(w_1,z_1)}^{(1)} P_{(w_2,z_2)}^{(1)} ... P_{(w_{pN},z_{pN})}^{(1)}\rho_0\right]$$

(10)

where $A_w$ and $A_z$ are antisymmetrization operators acting respectively on hole coordinates *w*'s and electron coordinates *z*'s and where $\rho_0$ is given by Eq. (4). The presence in Eq. (10) of (i) $\rho_0$ and (ii) projectors associated with angular momentum *L=1* respectively ensures that $\rho_{\beta_0}$ takes significant values only for coordinate configurations in which (i) each coordinate (electron or hole coordinate) occupies a surface of area approximately equal to $2\pi$ and (ii) a hole coordinate can be found in the vicinity of *every* electron coordinate, i.e. at a distance of the order of the magnetic length. The integral implicit in the projectors of Eq. (10) can be evaluated using the identity [9]

$$\int\prod_{i=1}^{n}\left[d^2z_i e^{-\frac{|z_i|^2+z_i^*\zeta}{2}}\right]P_k^*(z_1,...,z_n)Q_l(z_1,...,z_n)f(z_1,...,z_n) = f(\zeta,...,\zeta)\int\prod_{i=1}^{n}\left[d^2z_i e^{-\frac{|z_i|^2}{2}}\right]P_k^*(z_1,...,z_n)Q_l(z_1,...,z_n)$$

(11)

where *f* is a polynomial and where $P_k$ and $Q_l$ are homogeneous polynomials of total degree *k* and *l* with $k \leq l$ and are invariant under global translations of their variables. This leads to

$$\rho_{\beta_0} \propto A_w A_z \left[\prod_{i=1}^{pN} e^{-\frac{|z_i|^2}{4}} \prod_{pN+1\leq j_1<j_2\leq hN}(w_{j_1}-w_{j_2})\prod_{j=1}^{hN} e^{-\frac{|w_j|^2}{4}} \int\prod_{i=1}^{pN}\left(d^2\zeta_i(w_i-z_i)e^{\frac{w_i+z_i}{2}\zeta_i^*-\frac{|\zeta_i|^2}{2}}\right)\prod_{1\leq i_1<i_2\leq pN}(\zeta_{i_1}-\zeta_{i_2})^4\prod_{j=pN+1}^{hN}\prod_{i=1}^{pN}(\zeta_i-w_j)^2\right]$$

(12)

We now argue that the variable $\zeta_i$, which plays the role in Eq. (12) of guiding center for coordinates $z_i$ and $w_i$, is the coordinate of a composite Fermion. In order to see that, let us evaluate the partition function of the electron system as a path integral involving coordinates $\zeta$'s. The partition function for inverse temperature $\beta=M\beta_0$, with *M* integer and even, is first expressed as

$$Z(M\beta_0) = \int d^{2hN}W^{(1)} d^{2pN}Z^{(2)} d^{2hN}W^{(3)} ... d^{2pN}Z^{(M)}$$
$$\times \langle W^{(1)}|e^{-\beta_0 \bar{V}}|Z^{(2)}\rangle\langle Z^{(2)}|e^{-\beta_0 \bar{V}}|W^{(3)}\rangle\langle W^{(3)}|...|Z^{(M)}\rangle\langle Z^{(M)}|e^{-\beta_0 \bar{V}}|W^{(1)}\rangle$$
(13)

where $W^{(1)} \equiv (w_1^{(1)},...,w_{hN}^{(1)})$ and $Z^{(2)} \equiv (z_1^{(2)},...,z_{pN}^{(2)})$. We now use Eq. (12) to explicit the density matrix elements in Eq. (13). Neglecting permutations in $A_w$'s which connect a coordinate $w_i$ attached to a coordinate $\zeta_i$ in a given matrix element to a coordinate $w_{i'}$ unattached in the next matrix element and which lead to small contributions, one gets

$$Z(M\beta_0) \propto \int \prod_{i,k} d^2 \zeta_i^{(k)} R(\zeta_i^{(k)})$$
(14)

with function $R(\zeta_i^{(k)})$ given by

$$R(\zeta_i^{(k)}) = \sum_{m_i^{(k)}=0,1}(-1)^{\sum_{i,k} m_i^{(k)}} \prod_{l=1}^{M/2}\left[\prod_{1\le i_1 < i_2 \le pN}\left[(\zeta_{i_1}^{(2l-1)}-\zeta_{i_2}^{(2l-1)})(\zeta_{i_1}^{(2l)}-\zeta_{i_2}^{(2l)})^*\right]^4\right.$$
$$\times \quad \langle 0|\prod_i \{C(\zeta_i^{(2l-1)},m_i^{(2l-1)})\}\prod_i\{C^+(\zeta_i^{(2l)},m_i^{(2l)})\}|0\rangle$$
$$\times \quad \langle 0|\prod_i \{C(\zeta_i^{(2l-1)},1-m_i^{(2l-1)})\}\prod_i\{C^+(\zeta_i^{(2l-2)},1-m_i^{(2l-2)})\}|0\rangle$$
$$\times \quad \left.\langle 1,(h-p)N|\prod_i\{S^+_{\zeta_i^{(2l-1)}}{}^2\}\prod_i\{S_{\zeta_i^{(2l-2)}}{}^2\}|1,(h-p)N\rangle\right] ,$$
(15)

where indices $i$ and $k$ take values $i=1,...,pN$ and $k=1,...,M$, where $\zeta_i^{(0)} \equiv \zeta_i^{(M)}$, $m_i^{(0)} \equiv m_i^{(M)}$, where

$$C^+(\zeta,m) = \left(2\frac{\partial}{\partial\tilde{\zeta}}-\frac{\tilde{\zeta}^*}{2}\right)^m C^+_{\tilde{\zeta}}\Big|_{\tilde{\zeta}=\zeta}$$
(16)

is an electron creation operator and where $S_\zeta$ is the operator creating a quasihole [11], whose action on wave function $\varphi(z_1,...,z_n)$ is defined by

$$S_\zeta \varphi(z_1,...,z_n) = \prod_{i=1}^n (z_i^* - \zeta^*)\varphi(z_1,...,z_n).$$
(17)

The function $R(\zeta_i^{(k)})$ appearing in Eq. (14) can be interpreted as the exponential of minus the action of the composite Fermion system, evaluated on a given path $\zeta_i^{(k)}$.

We now take $M$ large, restrict to smooth paths, i.e. paths $\zeta_i^{(k)}$ with a smooth dependence on $k$, and evaluate the phase of $R(\zeta_i^{(k)})$ associated to such paths. Using the facts that (i) for most paths, the motion of $\zeta_i^{(k)}$ is equally distributed between odd $k$ to even $k$ steps and even $k$ to odd $k$ steps, (ii) for configurations $\zeta_i$, $i=1...n$, satisfying $|\zeta_i-\zeta_j|>(2m_i+1/2+(4m_i+1/4)^{1/2})^{1/2}+(2m_j+1/2+(4m_j+1/4)^{1/2})^{1/2}$, $1\le i<j \le n$, and for small displacements $d\zeta_i$, one has

$$\arg\left(\langle 0|\prod_{i=n}^1\{C(\zeta_i,m_i)\}\prod_{i=1}^n\{C^+(\zeta_i+d\zeta_i,m_i)\}|0\rangle\right) = \sum_{1\le i<j\le n} d\vartheta_{ij} + \frac{1}{2}\sum_{i=1}^n |\zeta_i|^2 d\vartheta_i$$
(18)

where $\theta_i$ and $\theta_{ij}$ are respectively the argument of $\zeta_i$ and $\zeta_i-\zeta_j$, and (iii) the matrix element $<0|c(\zeta_i,m) c^+(\zeta_i+d\zeta_i,m')|0>$ is negligible for $m\ne m'$, we obtain

$$\arg(R(\zeta_i(\tau))) = \int_0^\beta d\tau\left[\sum_{1\le i<j\le pN}\frac{d\vartheta_{ij}}{d\tau} + \frac{1}{2}\frac{h-p}{h+p}\sum_{i=1}^{pN}|\zeta_i|^2\frac{d\vartheta_i}{d\tau}\right]$$
(19)

where $\zeta_i(\tau)$ is the location of particle $i$ at imaginary time $\tau=k\beta_0$, defined by $\zeta_i(\tau) \equiv \zeta_i^{(k=\text{int}(\tau/\beta_0))}$. The phase change given by Eq. (19) is precisely what is expected for composite Fermions and we thus identify variables $\zeta_i(\tau)$ as composite Fermion coordinates. Hole coordinates $w_i$ attached to composite Fermion coordinates $\zeta_i$ in Eq. (12) contribute to the statistical phase change, which is described by the first term between brackets in Eq. (19), while unattached hole coordinates lead to the phase change associated with the effective magnetic field experienced by composite Fermions, which is described by the second term between brackets.

In conclusion, (i) this density matrix approach using both electron and hole coordinates allows to view the electron-electron repulsion as an *attractive* electron-hole interaction and (ii) near *ν=1/2*, composite Fermions can be regarded as clusters made up of an electron and a hole stabilized by the short ranged electron-hole attraction. Similarly, near *ν=1/4*, composite Fermions can be regarded as clusters made up of an electron and three holes.

Let us finally discuss the implications of this picture of composite Fermions for the incompressible *ν=5/2* state. In absence of mixing between Landau levels, the system of electrons filling half of the *n=1* Landau level and subject to Coulomb interaction can be mapped onto a system of electrons filling half of the *lowest* Landau level and subject to an electron-electron interaction differing from Coulomb at short distance [12]. The existence of the *ν=5/2* state is attributed [13] to the pairing of composite Fermions. By using the above–mentioned mapping to the lowest Landau level, this pairing can in turn be attributed [12] to the particular form taken by the pseudo-potential coefficients describing the electron-electron interaction in the equivalent system of electrons confined to the lowest Landau level. Since both pairing of composite Fermions and formation of composite Fermions as electron-hole pairs result dynamically from the inter-particle interaction, they should be described on the same footing in this density matrix approach. In other words, the incompressible *ν=5/2* state can be regarded as resulting from the formation of clusters containing each two electrons and two holes.

## 3. COMPOSITE BOSONS

We now present a formulation of the composite Boson approach of incompressible Fractional Quantum Hall states based on the density matrix formalism developed above. We make the case that the density matrix formalism developed here allows a direct formulation of the composite Boson approach of incompressible states for filling factors *ν=p/q* with arbitrary numerator *p≥1*, in contrast to the formulation based on singular gauge transformation in which *ν=p/q* states are constructed iteratively.

Following Ref. [9], we assume that, for some value $\beta_0$ of the inverse temperature, the density matrix for *ν=p/q* can be written

$$\rho_{\beta_0} \propto A_w A_z \left[ P_{(w_1,...,w_h,z_1,...,z_p)} P_{(w_{h+1},...,w_{2h},z_{p+1},...,z_{2p})} ... P_{(w_{(N-1)h+1},...,w_{Nh},z_{(N-1)p+1},...,z_{Np})} \rho_0 \right] \quad (20)$$

where *h=q-p* and where $P_{(w_1,...,w_h,z_1,...,z_p)}$ is the projection operator associated to wave function

$$\Psi^{(\zeta)}(w_1,...,w_h,z_1,...,z_p) = Q(w_1^*,...,w_h^*,z_1^*,...,z_p^*) \prod_{j=1}^{h} e^{\frac{w_j^* \zeta}{2} - \frac{|\zeta|^2 + |w_j|^2}{4}} \prod_{i=1}^{p} e^{\frac{z_i^* \zeta}{2} - \frac{|\zeta|^2 + |z_i|^2}{4}}, \quad (21)$$

where *Q* is a polynomial of total degree *q(q-1)/2* which is invariant under global translation of its variables. As in Eq. (9), the action of $P_{(w_1,...,w_h,z_1,...,z_p)}$ on function $\rho_0$ is defined by priming variables $w_1,...,w_h,z_1,...,z_p$ appearing in $\rho_0(w_1,...,w_{hN},z_1,...,z_{pN})$, then multiplying by the kernel

$$K(w_1,...,w_h,z_1,...,z_p,w'_1,...,w'_h,z'_1,...,z'_p) = \pi^{-1} \int d^2\zeta \, \Psi^{(\zeta)*}(w_1,...,w_h,z_1,...,z_p) \Psi^{(\zeta)}(w'_1,...,w'_h,z'_1,...,z'_p) \quad (22)$$

and finally by integrating over primed coordinates *w'₁,…,w'ₕ,z'₁,…,z'ₚ*. The presence of (i) $\rho_0$ and (ii) projectors in Eq. (20) respectively ensures that $\rho_{\beta_0}$ is non-negligible only for configurations *(w₁,…,w_{hN},z₁,…,z_{pN})* in which (i) each coordinate occupies a surface of area approximately equal to *2π* and in which (ii) coordinates can be grouped into clusters containing *h* hole coordinates and *p* electron coordinates ( denoted hereafter as *(p,h)* clusters ).

In a manner analogous to that described by Eqs. (13-19), the partition function of the electron system can be expressed as a path integral over paths $\zeta_i(\tau), i=1,…,N$, which specify the location of clusters *i=1,…,N* as a function of imaginary time *τ*. The phase corresponding to a given path is found to be equal to *(-1)^{ph{P}}*, where *{P}* is the parity of the permutation taking place among clusters between imaginary time *τ=0* and *τ=β*. Thus, paths contribute constructively to the partition function only if *ph* is even. In the latter case, paths involving large ring exchange can significantly contribute to the partition function at low temperature, leading to superfluidity [14]. Expressed in other words, if *ph* is even, *(p,h)* clusters obey Bose statistics and can Bose-condense at low temperature. Note however that these clusters do not maintain their integrity throughout imaginary time in the sense that among the permutations present in the antisymmetrization operators $A_w$ and $A_z$ of Eq. (20), those leading to the exchange of electron coordinates or hole coordinates between different clusters when moving from one imaginary time step to the next one can significantly contribute to the partition function.

This formulation of the composite Boson approach allows to present a very simple explanation for the fact that all incompressible states found in the lowest Landau level appear at filling fraction *ν=p/q* with odd denominator. Indeed, in the generic case of a short ranged interaction between electrons confined to the lowest Landau level, it is likely that *(p/2,h/2)* clusters will be energetically preferred to *(p,h)* clusters if *p* and *h* are both even. When the cluster cannot be halved anymore, *p* and *h* cannot be both even. This latter condition, together with the condition that *ph* be even, implies that *q=p+h* be odd. This argument fails in the case of the *ν=5/2* state, because clusters with *p=h=2* are stabilized by the particular form taken by the electron-electron interaction at short distance when projected into the *n=1* Landau level.

Also, a simple and generic description can be presented for elementary excitations of incompressible states using this formulation of the composite Boson approach. In the picture of an incompressible *ν=p/q* state in terms of *(p,h)* clusters, a neutral excitation corresponds to the breaking of a *(p,h)* cluster into a *(p',h')* cluster and a *(p",h")* cluster, with *p'+p"=p* and *h'+h"=h*. Choosing *p'* and *h'* such that *p'h-ph'=1* in order to minimize variations in charge density [9] and taking into account the fact that each electron or hole occupy a surface with area equal to *2π*, one concludes directly that each smaller cluster is associated to a charge defect *±1/q* in units of the electron's charge, i.e. is associated to a quasiparticle.

## 4. ACKNOWLEDGEMENTS

We would like to thank C. Renault for help in preparing this manuscript and R. Morf for helpful comments.